%% file: main.tex
\documentclass{sig-alternate}
\input{macros}
\usepackage{booktabs}

\newcommand{\weight}{w}
\usepackage{url}
 \usepackage{multirow}
 \usepackage{tikz}  
\usepackage{etoolbox}
\makeatletter
\patchcmd{\maketitle}{\@copyrightspace}{}{}{}
\makeatother

\usepackage{fancyhdr}
\renewcommand{\headheight}{0.6in}
\begin{document}

\title{iWinRNFL: A Simple, Interpretable \& Well-Calibrated In-Game Win Probability Model for NFL} 

\numberofauthors{1} 
%
\author{
%
%
\alignauthor
Konstantinos Pelechrinis\\
        \affaddr{School of Computing and Information}\\
       \affaddr{University of Pittsburgh}\\
       \email{kpele@pitt.edu}
}

\date{}
\maketitle

\begin{abstract}
During the last few {\em sports seasons} a lot of discussion has been generated for the several, high-profile, ``comebacks'' that were observed in almost all sports.  
The Cavaliers won the championship after being down 3-1 in the 2016 NBA finals' series against the Golden State Warriors, which was exactly the case for Chicago Cubs and the World Series.  
The Patriots won the Super Bowl in 2016 even though they were trailing by 25 points late in the third quarter, while FC Barcelona in the top-16 round of the 2016-17 Champions League scored 3 goals during the last 7 minutes of the game (including stoppage time) against PSG to advance in the tournament. 
This has brought the robustness and accuracy of the various probabilistic prediction models under high scrutiny.  
Many of these models are proprietary, which makes it hard to evaluate. 
In this paper, we build a simple and open, yet robust and well-calibrated, \underline{i}n-game probability model for predicting the \underline{win}ne\underline{r} in an  \underline{NFL} ({\method}) game. 
In particular, we build a logistic regression model that utilizes a set of 10 variables to predict the running win probability for the home team.  
We train our model using detailed play-by-play data from the last 7 NFL seasons obtained through the league's API. 
Our results indicate that in 75\% of the cases {\method} provides an accurate winner projection, as compared to a 63\% accuracy of a baseline pre-game win probability model. 
Most importantly the probabilities that {\method} provides are well-calibrated. 
Finally, we have also evaluated more complex, non-linear, models using the same set of features, without any significant improvement in performance.  
\end{abstract}
 
\section{Introduction}
\label{sec:intro}

In-game win probability models provide the likelihood that a certain team will win a game given the current state of the game.  
These models have become very popular during the last few years, mainly because they can provide the backbone for in-game decisions as well as, play call and personnel evaluations. 
Furthermore, they can potentially improve the viewing experience of the fans.  
Among other events Super Bowl 51 sparked a lot of discussion around the validity and accuracy of these models \cite{ringer17}. 
During Super Bowl 51, late in the third quarter the probability assigned by some of these models to New England Patriots to win was less than 0.1\% or in other words a 1 in 1,000 Super Bowls comeback \cite{statsbylopez}.  
Clearly the result of the game was not the one projected at that point and hence, critiques of these models appeared.  
Of course, this is a single observation and a large part of the discussion was generated because of the high profile of the game and selection bias. 
Nevertheless, designing and evaluating in-game win probability models is in general important if we are to rely on them for on-field decision making and for evaluating teams, players and coaches. 
Furthermore, the proprietary nature of many of these models makes it imperative to develop ones that are open and can be further inspected, reviewed and improved by the community, while simpler and interpretable models are always preferable over more complicated and hard to interpret models (without sacrificing quality). 
For example, \cite{lock2014using} is a very similar model to {\method}, with similar performance. 
However, it makes use of a more complex model, that is, ensemble learning methods, that might be harder to be directly interpreted in a way similar to a linear model.  
We would like to emphasize that our study does not aim at discrediting existing win probability models, but rather exploring the ability of simple and interpretable models to achieve similar performance and identify when they can {\em fail}. 

Therefore, in this paper we present the design of {\method}, an in-game win probability model for NFL.  
Our model was trained using NFL play-by-play data from 7 NFL seasons between 2009-2015, and while {\method} is simple - at its core is based on a generalized linear model - our evaluations show that it is well-calibrated over the whole range of probabilities.  
The real {\bf robustness} question in these type of models is how well the predicted probabilities capture the actual winning chances of a team at a given point in the game.  
In other words, what is the {\em reliability} of our predictions; i.e., what fraction of the in-game instances where a team was given x\% probability of winning actually ended up with this team winning?  
Ideally, we would like this fraction to be also x\%.  
For instance, the fact that New England Patriots won Super Bowl 51 even though they were given just 0.1\% probability to do so at the end of the third quarter, is not a {\em failure of the math}.  
In fact, this number tells us that roughly for every 1,000 times that a team is in the same position there will be 1 instance (on expectation) where the trailing team will win.  
Of course, the order with which we observe this instance is arbitrary, that is, it can be the first time we ever observe this game setting (i.e., a team trailing with 25 points by the end of the third quarter), which further intensifies the opinion that {\em math failed}. 
Our evaluations show that this is the case for {\method} over the whole range of probabilities.  

\begin{figure}[t]
\begin{center}
\includegraphics[scale=0.45]{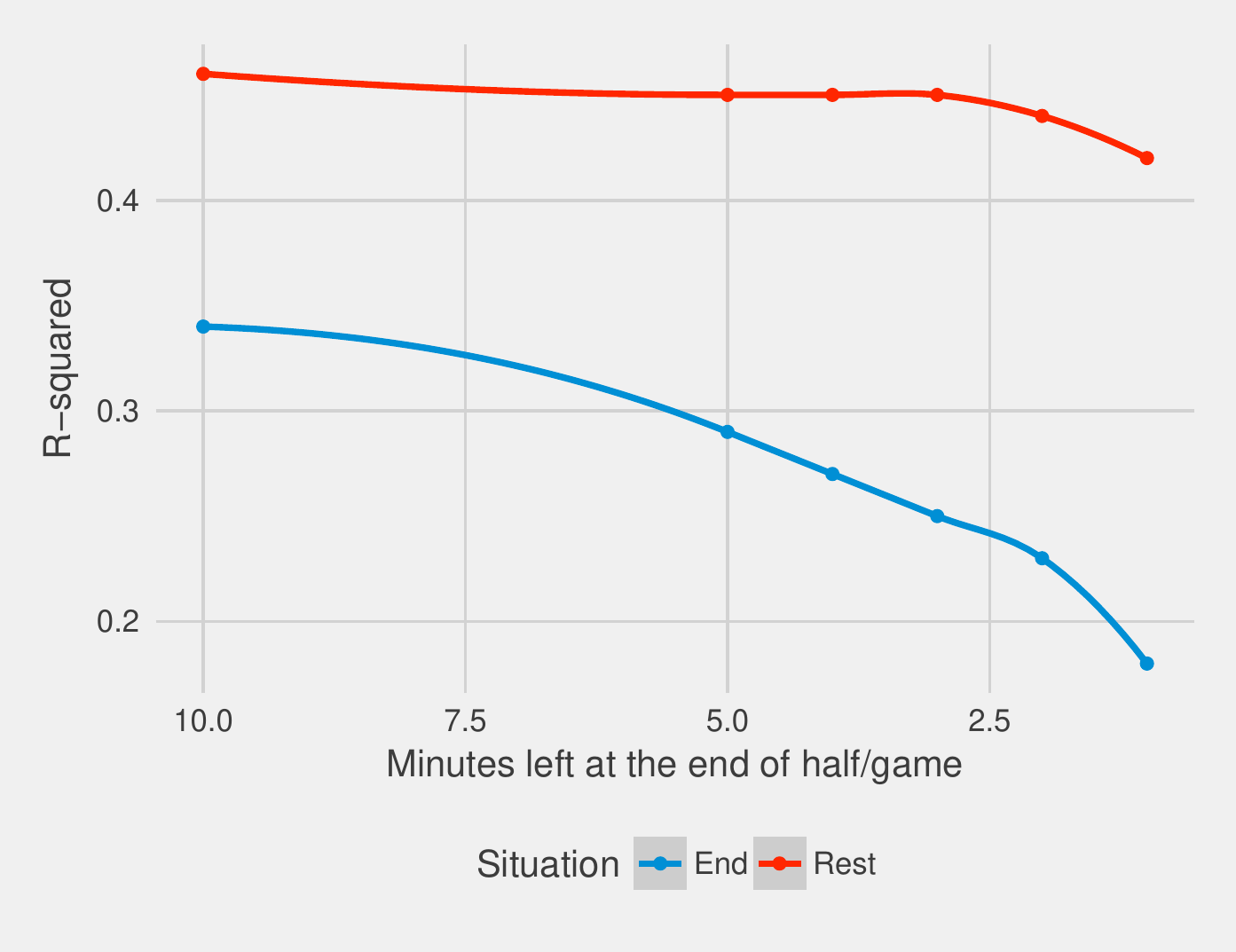}
 \caption{A linear hypothesis between time and yard length of a drive cannot explain a lot of the variance towards drives at the end of the half/game. }
 \label{fig:non-linear}
\end{center}
\end{figure}

Furthermore, we evaluate more complex, non-linear, models, and in particular a Bayesian classifier and a neural network, using the same set of features. 
The reason for examining more complex models are the boundary effects from the finite duration of a game that can create several non-linearities \cite{winston2012mathletics} and these are the cases that simple, linear modes can fail. 
For example, when we examine the relationship between a drive's time length and a drive's yard length, a linear model between these two variables explains approximately 42\% of the variance. 
However, when focusing at the end of the half/game, this linear model explains much less of this variance. 
In fact, the closer we get to the end of the half/game, the lower the quality of the linear model as we can see in Figure \ref{fig:non-linear}.  
In this figure, we present the variance explained ($R^2$) by a linear model between the two variables for two different types of drives, namely, drives that started $\tau$ minutes before the end of the half/game and drives that started outside of this window. 
As we can see, the closer we get to the end of the half/game, the less variance this linear relationship can explain, serving as evidence that indeed there can be much more severe non-linearities towards the end of the second and fourth quarter - as compared to earlier in the first and second half. 
However, as we will see in our evaluations, overall there are not significant performance improvements from these non-linear models over {\method}, despite these possibly un-modeled non-linear factors contributing to win probability! 

The rest of the paper is organized as follows: 
Section \ref{sec:background} briefly presents background on (in-game) win probability models for NFL.  
Section \ref{sec:model} presents our model(s), while Section \ref{sec:evaluation} presents the model evaluations.  
Finally, Section \ref{sec:conclusions} concludes our work. 

\section{Background}
\label{sec:background}

Predicting the outcome of a sports game - and in particular American football in our case - has been of interest for several decades now. 
For example, \cite{stern91} used data from the 1981, 1983 and 1984 NFL seasons and found that the distribution of the win margin is normal with mean equal to the pregame point spread and standard deviation a little less than 14 points.  
He then used this observation to estimate the probability distribution of the number of games won by a team.  
The most influential work in the space, Win Probability Added, has been developed by \cite{wpa} who uses various variables such as field position, down etc. to predict the win probability added after every play.  
This work forms the basis for ESPN's prediction engine, which uses an ensemble of machine learning models.   
Inspired by the early work and observations from Stern, Winston developed an in-game win-probability model \cite{winston2012mathletics}, which was further adjusted from Pro-Football Reference to form their P-F-R win probability model using the notion of expected points \cite{pfrmodel}.  
More recently, \cite{lock2014using} provided a random forest model for in-game win probability, while \cite{gambletron} created a system that uses real-time data from betting markets to estimate win probabilities. 
As alluded to above \cite{lock2014using} used very similar covariates with {\method} and the performance of the two models is very similar. 
One of the reasons the authors used random forests is the ability of the model to account for non-linear interactions between the covariates. 
Our work shows that the improvements (if at all) over simpler and more interpretable, (generalized) linear models, do not justify the use of complex models. 

At the wake of Super Bowl 51 \cite{rosenheck17} developed a simulation system that considers the strength of the offensive and defensive units, the current score differentials and the field position and simulates the rest of the game several times to obtain the current win probability.  
This approach is computational expensive and while, the goal of Rosenheck was to retrospectively simulate the outcome of Super Bowl 51 under the assumption that the overtime rules for the NFL are similar to those of college football, its applicability in its current form is mainly for post-game analysis.  
Finally, similar in-game win probability models exist for other sports (e.g., \cite{buttrey2011estimating,stern1994brownian}). 

One of the main reasons we introduce {\method} despite the presence of several in-game win probability models is the fact that the majority of them are hard to reproduce either because they are proprietary or because they are not described in great detail.  
Furthermore, an older model might not be applicable anymore ``as-is'' given the changes the game undergoes over the years (e.g., offenses rely much more on the passing game today as compared to a decade ago, changes in the rules etc.). 
Our main objective with this work is to provide an {\em open} and fully {\em reproducible} win probability model for the NFL\footnote{Source code and data will be made publicly available.}. 


\section{The iWinRNFL model}
\label{sec:model}

In this section we are going to present the data we used to develop our in-game probability model as well as the design details of {\method}. 

\begin{figure*}[t]
\begin{center}
\includegraphics[scale=0.3]{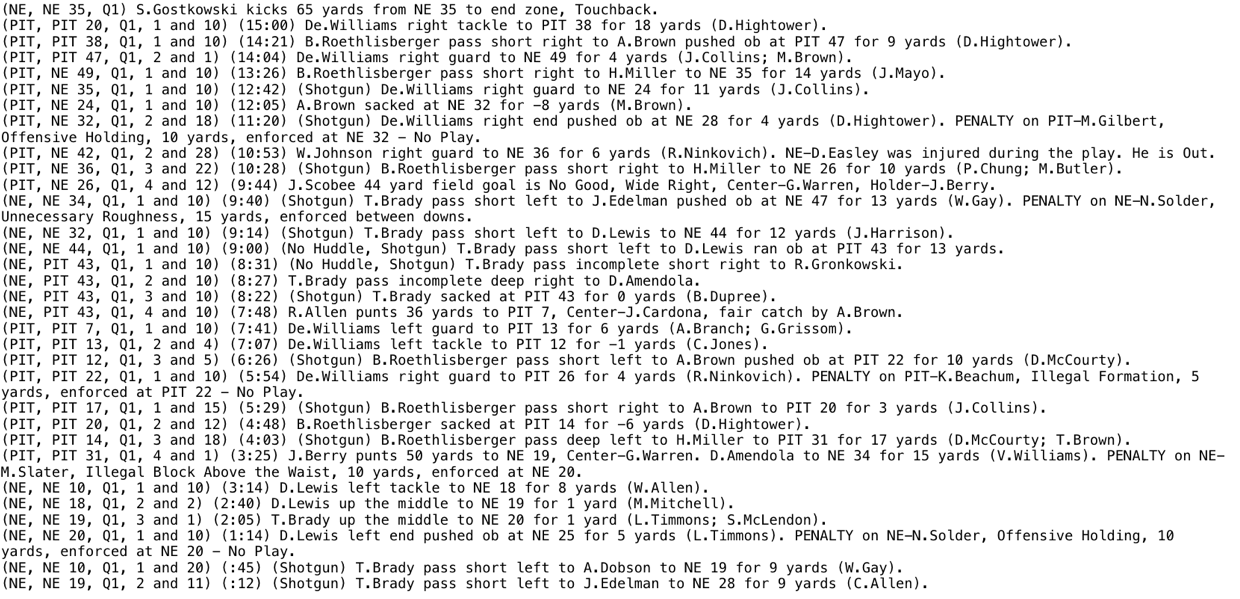}
 \caption{Through Python's {\tt nflgame} API we obtain a detailed log for every regular season NFL game between 2009-2016.}
 \label{fig:pbp}
\end{center}
\end{figure*}

{\bf Data: }In order to perform our analysis we utilize a dataset collected from NFL's Game Center for all the regular season games between the seasons 2009 and 2015. 
We access the data using the Python {\tt nflgame} \cite{nflgame}. 
The dataset includes detailed play-by-play information for every game that took place during these seasons. 
Figure \ref{fig:pbp} presents an illustrative sample of the game logs we obtain. 
This information is used to obtain the state of the game that will drive the design of {\method}. 
In total, we collected information for 1,792 regular season games and a total of 295,844 snaps/plays. 

{\bf Model: }
{\method} is based on a logistic regression model that calculates the probability of the home team winning given the current status of the game as: 

\begin{equation}
\Pr(H=1| \mathbf{x})= \frac{\exp(\mathbf{\weight}^T\cdot\mathbf{x})}{1+\exp(\mathbf{\weight}^T\cdot\mathbf{x})}
\label{eq:reg}
\end{equation}
where $H$ is the dependent random variable of our model representing whether the home team wins or not, $\mathbf{x}$ is the vector with the independent variables, while the coefficient vector $\mathbf{\weight}$ includes the weights for each independent variable and is estimated using the corresponding data.  



In order to describe the status of the game we use the following variables:

\begin{enumerate}
\item {\bf Ball Possession Team:} This binary feature captures whether the home or the visiting team has the ball possession
\item {\bf Score Differential:} This feature captures the current score differential (home - visiting)
\item {\bf Timeouts Remaining:} This feature is represented by two independent variables - one for the home and one for the away team - and they capture the number of timeouts remaining for each of the teams
\item {\bf Time Elapsed: } This feature captures the time elapsed since the beginning of the game
\item {\bf Down:} This feature represents the down of the team in possession
\item {\bf Field Position:} This feature captures the distance covered by the team in possession from their own yard line
\item {\bf Yards-to-go:} This variables represents the number of yards needed for a first down
\item {\bf Ball Possession Time: } This variable captures the time that the offensive unit of the home team is on the field 
\item {\bf Rating Differential: } This variable represents the difference in the ratings for the two teams (home - visiting)
\end{enumerate}

The last independent variable is representative of the strength difference between the two teams. 
The rating of each team $T$ represents how many points better (or worse) $T$ is compared to a league-average team. 
This rating differential {\em dictates} the win-probability at the beginning of the game, and its importance fades as the game progresses as we will see. 
Appendix A describes in detail how we obtain these ratings, as well as other feature alternatives for representing the strength difference. 
Furthermore, we have included in the model three interaction terms between the ball possession team variable, and (i) the down count, (ii) the yards-to-go, and (iii) the field position variables.
This is crucial in order to capture the correlation between these variables and the probability of the home team winning. 
More specifically, the interpretation of these three variables (down, yards-to-go and field position) is different depending on whether the home or visiting team possesses the ball and these interaction terms will allow the model to better distinguish between the two cases. 
Finally, we have added an interaction term between the time lapsed and (i) the team ratings differential and (ii) the score differential, in order to examine whether and how the importance of these covariates changes as the game progresses. 
Table \ref{tab:iwinrnfl} presents the coefficients of the logistic regression model of {\method} with standardized independent variables for better comparisons.

\begin{table}[ht]
\begin{center}
\def\sym#1{\ifmmode^{#1}\else\(^{#1}\)\fi}
\begin{tabular}{l*{1}{c}}
\toprule
                    &\multicolumn{1}{c}{Winner}\\
\midrule
Possession Team (H)         &      -0.88\sym{***}\\
Score Differential           &      1.41\sym{***}\\
Home Timeouts           &     0.06\sym{***}\\
Away Timeouts           &     -0.06\sym{***}\\
Ball Possession Time  &    - 0.46\sym{***}\\
Time Lapsed       &   0.43\sym{***}\\
Rating Differential         &       1.72\sym{***}\\
Down (1)               &   -0.39\sym{***} \\
Down (2)		& -0.29\sym{***} \\
Down (3) 		& -0.20\sym{***}\\
Down(4)    &	 -0.05\sym{***} \\   
Field Position            &  -0.41\sym{***} \\
Yards-to-go                &  0.07\sym{***}        \\
{\bf Interaction terms} & \\
Possession Team (H)$\cdot$ Down (1) & 0.65\sym{***} \\
Possession Team (H)$\cdot$ Down (2) & 0.47\sym{***} \\
Possession Team (H)$\cdot$ Down (3) & 0.30\sym{***} \\
Possession Team (H)$\cdot$ Down (4) & 0.08\sym{***} \\
Possession Team (H)$\cdot$ Field Position & 1.05\sym{***}\\
Possession Team (H)$\cdot$ Yards-to-go & -0.18\sym{***}\\
Time Lapsed $\cdot$ Rating Differential & -0.65\sym{***}\\
Time Lapsed $\cdot$ Score Differential & 2.88\sym{***}\\
\midrule
Observations        &      295,844         \\
\bottomrule
\multicolumn{2}{l}{\footnotesize \sym{.} \(p<0.1\), \sym{*} \(p<0.05\), \sym{**} \(p<0.01\), \sym{***} \(p<0.001\)}\\
\end{tabular}
\end{center}
\caption{Standardized logistic regression coefficients for {\method}.}
\label{tab:iwinrnfl}
\end{table}

\begin{figure}[t]
\begin{center}
\includegraphics[scale=0.5]{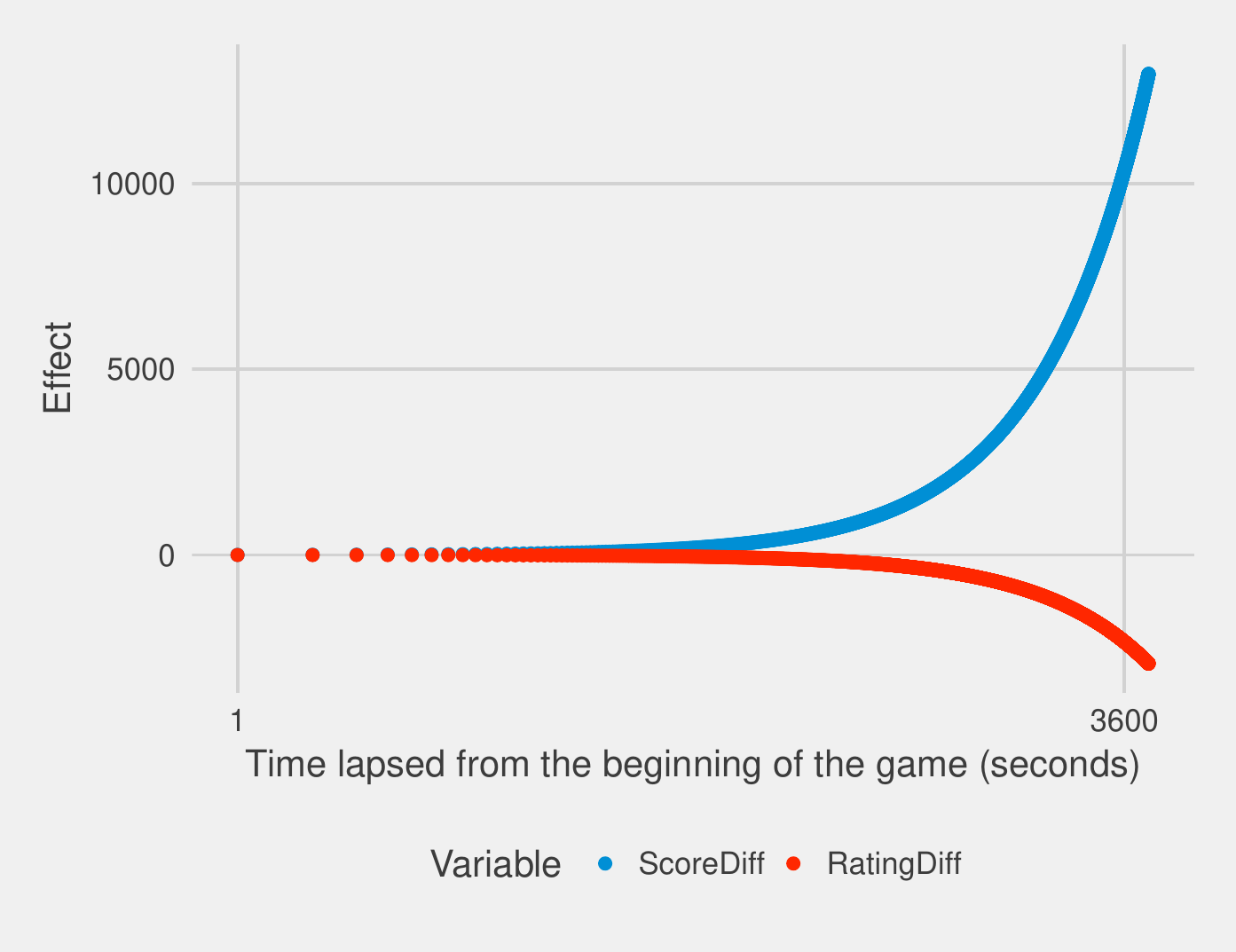}
 \caption{The effect of the team's strength differential decays as the game progresses, while that of the score differentials increases significantly ($x$-axis is in logarithmic scale for better visualization).}
 \label{fig:rtg-effect}
\end{center}
\end{figure}

As we can see, all of the factors considered are statistically significant for estimating the current win-probability for the home team. 
Particular emphasis should be given to the interaction terms. 
More specifically we see that - as one might have expected - having the ball on a first down provides a higher win probability as compared to a third or fourth down (for the same yards-to-go).  
Similarly, the probability of winning for the home team increases when its offensive unit is closer to the goal line (Field Position variable), while fewer yards to go for a first down are also associated with a larger win probability. 
Furthermore, an interesting point is the symmetric impact of the number of timeouts left for the home and visiting team.  
With regards to teams strength difference this appears to be crucial at the win probability at the beginning of the game, but its impact fades as time lapses. 
This is evident from the negative coefficient of the interaction term (Time Lapsed $\cdot$ Rating differential). 
In particular, the effect of the team rating differential on the win probability depends also on the time lapsed, and it is equal to $1.72-0.65\cdot(Time Lapsed)$. 
In other words, the coefficient for the rating differential (i.e., 1.7) captures only the effect of the rating differential at the very start of the game (i.e., Time Lapsed is 0). 
In contrast, as the game progresses the impact of the score differential on the (home) win probability increases, and it is equal to $1.41+2.88\cdot(Time Lapsed)$
Figure \ref{fig:rtg-effect} shows the declining impact of team rating differential as the game progresses in contrast with the increasing impact of current score differential  on win probability. 
Finally, it is worth noting that the intercept of the model is 0. 
One could have expected the intercept to capture the home field advantage \cite{kpele-plosone}, but the teams' rating differential has already included the home edge (see Appendix A).   
In the following section we provide a detailed performance evaluation of {\method}.

\section{Model Evaluation}
\label{sec:evaluation}

Before describing and computing our evaluation metrics, we will briefly describe two alternative models for estimating the win probability. 
In particular, we use the same features as above, but we evaluate two non-linear models, namely, a {\bf (naive) Bayesian classifier} and a {\bf feedforward neural network} (FNN). 
A naive Bayes classifiers computes the conditional probability of each class (home team win/loss in our case) for a data instance $k$ with independent variables {\bf x}=$(x_1,x_2,\dots,x_n)$, assuming conditional independence between the features given the true class $C_k$; i.e., $\Pr[x_i | x_1,x_2,\dots,x_n,C_k] = \Pr[x_i | C_k]$.  
Under this assumption the conditional probability for the class $C_k$ of data instance $k$ is given by: $\Pr[C_k | \mathbf{x}] = \Pr[C_k] \prod_{i=1}^n \Pr[x_i|C_k]$. 

\def\layersep{2.5cm}

\def\layerseptwo{5cm}

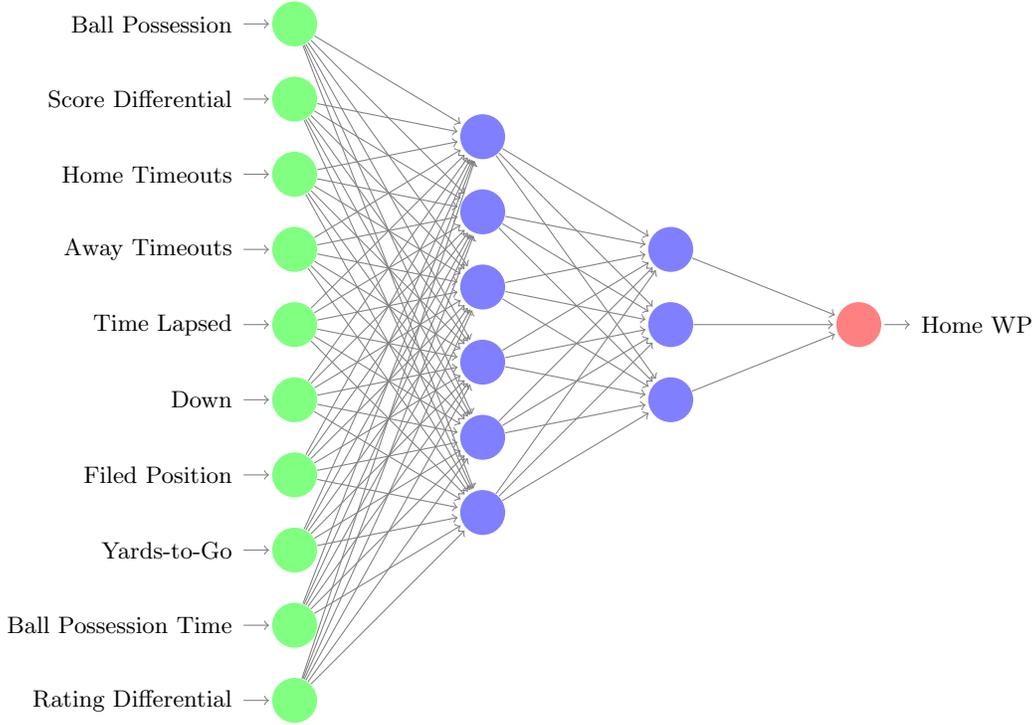
\begin{figure*}
\centering
\begin{tikzpicture}[shorten >=1pt,->,draw=black!50, node distance=\layersep]
    \tikzstyle{every pin edge}=[<-,shorten <=1pt]
    \tikzstyle{neuron}=[circle,fill=black!25,minimum size=17pt,inner sep=0pt]
    \tikzstyle{input neuron}=[neuron, fill=green!50];
    \tikzstyle{output neuron}=[neuron, fill=red!50];
    \tikzstyle{hidden neuron}=[neuron, fill=blue!50];
        \tikzstyle{hidden neuron2}=[neuron, fill=blue!50];
    \tikzstyle{annot} = [text width=4em, text centered]

        \node[input neuron, pin=left: Ball Possession] (I-1) at (0,5) {};
        \node[input neuron, pin=left: Score Differential] (I-2) at (0,4) {};
        \node[input neuron, pin=left: Home Timeouts] (I-3) at (0,3) {};
        \node[input neuron, pin=left: Away Timeouts] (I-4) at (0,2) {};
         \node[input neuron, pin=left: Time Lapsed] (I-5) at (0,1) {};
        \node[input neuron, pin=left: Down] (I-6) at (0,0) {};
         \node[input neuron, pin=left: Filed Position] (I-7) at (0,-1) {};
          \node[input neuron, pin=left: Yards-to-Go] (I-8) at (0,-2) {};
	 \node[input neuron, pin=left: Ball Possession Time] (I-9) at (0,-3) {};
	  \node[input neuron, pin=left: Rating Differential] (I-10) at (0,-4) {};
        \path[yshift=0.5cm]
            node[hidden neuron] (H-1) at (\layersep,3 cm) {};
            \path[yshift=0.5cm]
             node[hidden neuron] (H-2) at (\layersep,2 cm) {};
              \path[yshift=0.5cm]
            node[hidden neuron] (H-3) at (\layersep,1 cm) {};
            \path[yshift=0.5cm]
            node[hidden neuron] (H-4) at (\layersep,0 cm) {};
            \path[yshift=0.5cm]
             node[hidden neuron] (H-5) at (\layersep,-1 cm) {};
              \path[yshift=0.5cm]
            node[hidden neuron] (H-6) at (\layersep,-2 cm) {};

        \path[yshift=0.5cm]
            node[hidden neuron2] (HH-1) at (\layerseptwo,1.5 cm) {};
             \path[yshift=0.5cm]
            node[hidden neuron2] (HH-2) at (\layerseptwo,0.5 cm) {};
            \path[yshift=0.5cm]
            node[hidden neuron2] (HH-3) at (\layerseptwo,-0.5 cm) {};

    \node[output neuron,pin={[pin edge={->}]right:Home WP}, right of=HH-2] (O) {};

    \foreach \source in {1,...,10}
        \foreach \dest in {1,...,6}
            \path (I-\source) edge (H-\dest);

    \foreach \source in {1,...,6}
    	  \foreach \dest in {1,...,3}
	    \path (H-\source) edge (HH-\dest);
       
  \foreach \source in {1,...,3}
        \path (HH-\source) edge (O);

\end{tikzpicture}
\caption{The Feedforward Neural Network we used for the win probability includes two hidden layers (blue nodes).} \label{fig:fnn}
\end{figure*}

We also build a win probability model using a feedforward neural network with 2 hidden layers (Figure \ref{fig:fnn}). 
The first hidden layer has a size of 6 nodes, while the second hidden layer has a size of 3 nodes. 
While the goal of our work is not to identify the {\em optimal} architecture for the neural network, we have experimented with a different numbers and sizes of hidden layers and this architecture provided us with the best performance on a validation set\footnote{The performance of other architectures was not much different.}.  

We begin by evaluating how well the output probabilities of {\method} follow what happens in reality.  
When a team is given a 70\% probability of winning at a given state of the game, this essentially means that if the game was played from that state onwards 1,000 times, the team is expected to win approximately 700 of them.  
Of course, it is clear that we cannot have the game played more than once so one way to evaluate the probabilities of our model is to consider all the instances where the model provided a probability for team winning $x\%$ and calculate the fraction of instances that ended up in a win for this team.  
Ideally we would expect this fraction to be $x\%$ as well. 
This is exactly the definition of the reliability curve of a probability model. 

\begin{figure}[t]
\begin{center}
\includegraphics[scale=0.3]{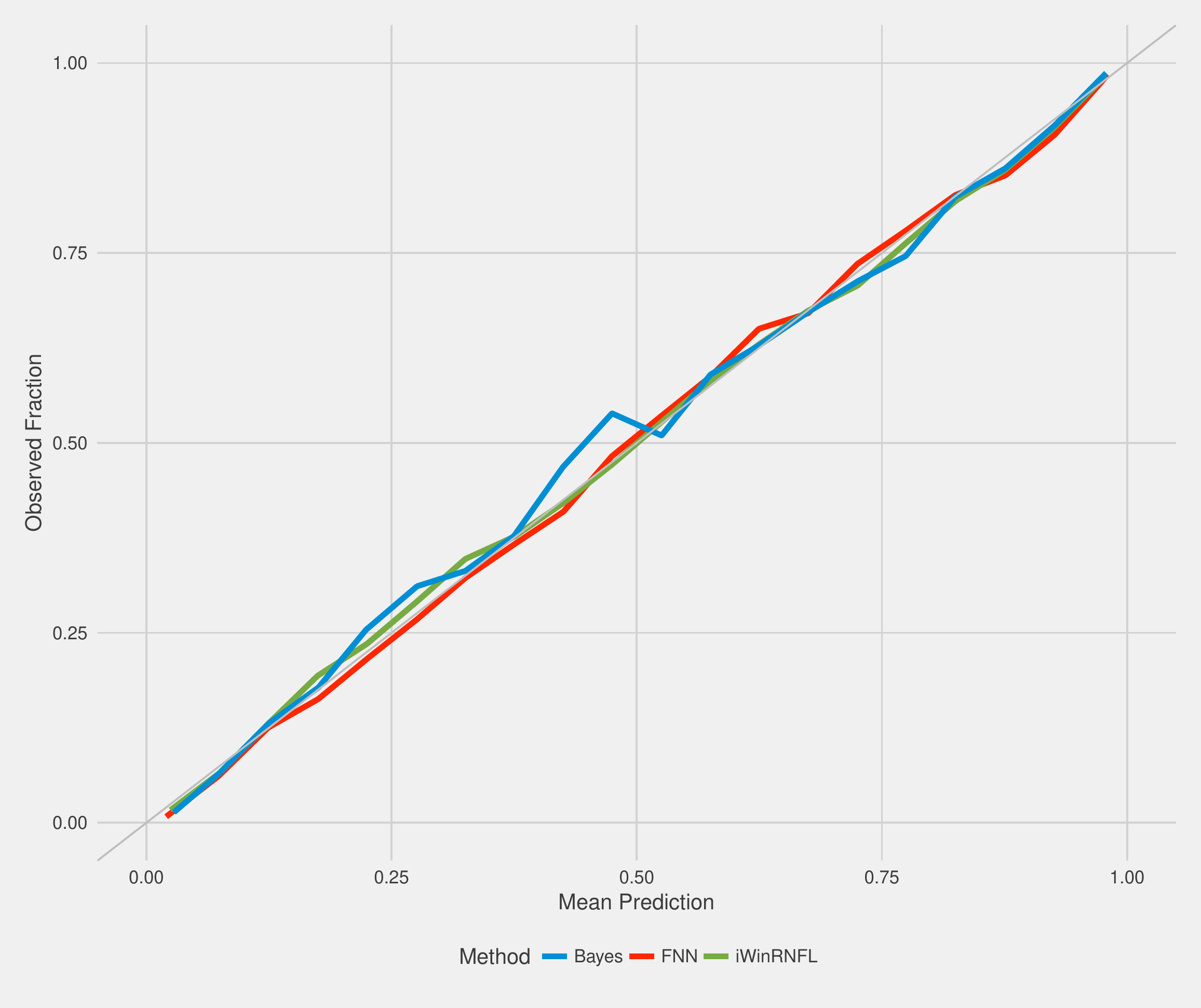}
 \caption{{\method} is as well-calibrated as the non-linear models examined over the whole range of probabilities.}
 \label{fig:fullwp}
\end{center}
\end{figure}

In order to obtain these results we split our data in a training and test set in a 70-30\% proportion respectively.    
Figure \ref{fig:fullwp} presents the results on our test set, where we used bins of a 0.05 probability range. 
In particular, as we can see the predicted probabilities match very well with the actual outcome of these instances.  
The fitted line ($R^2 = 0.998$) has a slope of 0.98 (with a 95\% confidence interval of [0.97,1.01]), while the intercept is 0.008 (with a 95\% confidence interval [-0.001, 0.02]). 
Simply put the line is for all practical purposes the $y=x$ line, which translates to a fairly consistent and accurate win probability. 

We have also calculated the accuracy of our binary predictions on the test set.  
In particular, the home team is projected to win if $\Pr(H=1| \mathbf{x})>0.5$. 
The accuracy of {\method} is equal to 76.5\%.   
The accuracy of the other two models examined is very similar, with naive Bayes exhibiting a 75\% accuracy, while the feedforward neural network has an accuracy of 76.3\%. 
 
Another metric that has been traditionally used in the literature to evaluate the performance of a probabilistic prediction is the Brier score $\brier$ \cite{brier1950verification}.  
In the case of a binary probabilistic prediction the Brier score is calculated as: 

\begin{equation}
\brier = \dfrac{1}{N}\sum_{i=1}^N (\pi_i-y_i)^2
\label{eq:brier}
\end{equation}
where $N$ is the number of observations, $\pi_i$ is the probability assigned to instance $i$ being equal to 1 and $y_i$ is the actual (binary) value of instance $i$.  
Brier scores takes values between 0 and 1 and does not evaluate the accuracy of the predicted probabilities but rather the calibration of these probabilities, that is, the level of certainty they provide.  
The lower the value of $\brier$ the better the model performs in terms of calibrated predictions. 
{\method} exhibits a Brier score $\brier$ of 0.158.  
Typically the Brier score of a model is compared to a baseline value $\brier_{base}$ obtained from a {\em climatology} model \cite{mason2004using}. 
A climatology model assigns the same probability to every observation (that is, home team win in our case), which is equal to the fraction of positive labels in the whole dataset.  
Hence, in our case the climatology model assigns a probability of 0.57 to each observation, since 57\% of the instances in the dataset resulted to a home team win. 
The Brier score for this reference model is $\brier_{base}=0.26$, which is obviously of lower quality as compared to our model. 
Both the naive Bayes and the FNN models exhibit similar performance with {\method} with Brier scores of 0.163 and 0.156 respectively. 

\begin{figure}[t]
\begin{center}
\includegraphics[scale=0.3,angle=270]{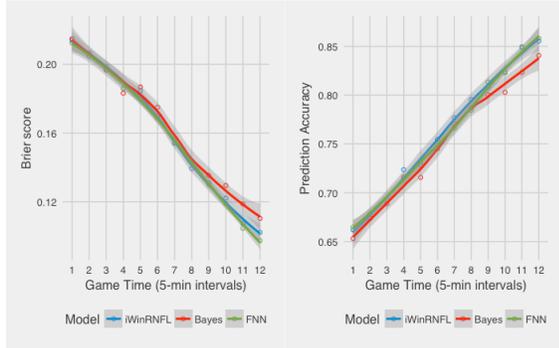}
 \caption{All models' performance improves later in the game, while FNN provides only incremental improvements over {\method} towards the end of the half/game.}
 \label{fig:performance-time}
\end{center}
\end{figure}

As alluded to above one of the reasons we examined the performance of more complex, non-linear, models is the fact that the finite duration of the half/game can introduce non-linearities that are not possible to be captured by {\method}. 
Therefore, apart from the overall performance of the different models, we would also like to examine performance of {\method} as a function of the time elapsed from the beginning of the game and compare it with the naive Bayes and FNN. 
More specifically, we consider 5-minutes intervals during the game; e.g., the first interval includes predictions that took place during the first 5 minutes of the game, while interval 7 includes predictions that took place during the fist five minutes of the third quarter. 
Figure \ref{fig:performance-time} depicts our results. 
As we can see the performance of all models is very close to each other and improves as the game progresses, as one might have expected. 
Furthermore, the prediction accuracy during the beginning of the game is very close to the state-of-the-art prediction accuracy of pre-game win probability models \cite{kpele-plosone}. 
This is again expected since at the beginning of the game the teams' rating differential is important, while as the game progresses the importance of this covariate reduces as we saw earlier (see Figure \ref{fig:rtg-effect}).  
More importantly, we see some improvement of FNN over {\method} particularly with regards to the Brier score and during the end of the game (intervals 11 and 12), but this improvements is marginal and cannot practically justify the use of a more complex model over a simple (interpretable) general linear model. 

{\bf Anecdote {\em Evaluation}: }
As mentioned in the introduction, one of the motivating events for {\method} has been Super Bowl 51. 
Super Bowl 51 has been labeled as the biggest comeback in the history of Super Bowl.  
Late in the third quarter New England Patriots were trailing by 25 points and Pro Football Reference was giving Patriots a 1:1000 chance of winning the Super Bowl, while ESPN a 1:500 chance \cite{statsbylopez}. 
PFR considers this comeback a once in a millennium comeback. 
While this can still be the case\footnote{As mentioned earlier 1:x chances does not mean that we observe x {\em failures} first and then the one {\em success}.}, in retrospect Patriots' win highlights that these models might be too optimistic and confident to the team ahead.  
On the contrary, the lowest probability during the game assigned to the Patriots by {\method} for winning the Super Bowl was 2.1\% or about 1:50.  
We would like to emphasize here that the above does not mean that {\method} is ``{\em better}'' than other win probability models.  
However, it is a simple and most importantly transparent model that assigns win probabilities in a conservative (i.e., avoids ``over-reacting''), yet accurate and well-calibrated way. 

\begin{figure}[t]
\begin{center}
\includegraphics[scale=0.4]{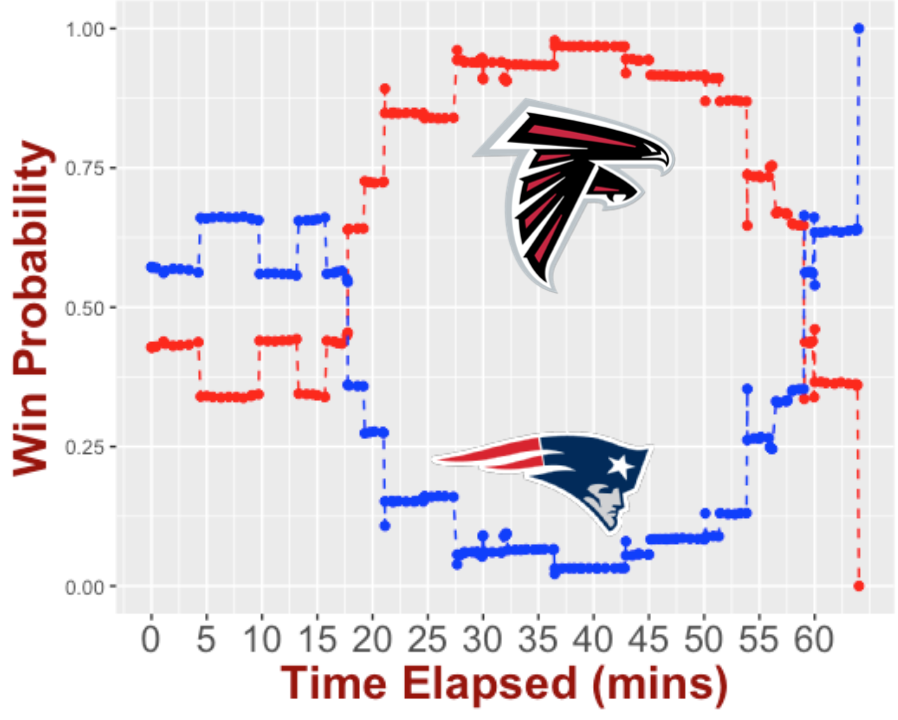}
 \caption{The lowest in-game win probability assigned to the Patriots by {\method} during Super Bowl 51 was 2.1\%, i.e., 1 in 50 chances. During the OT the model does not perform very accurately due to the sparsity of the relevant data.}
 \label{fig:sb51}
\end{center}
\end{figure}

\section{Conclusions}
\label{sec:conclusions}

In this paper, motivated by several recent comebacks that seemed improbable at the moment, we designed {\method}, a simple and open generalized linear model for in-game win probabilities in NFL. 
{\method} uses 10 independent variables to assess the win probability of a team for any given state of the game. 
Our evaluations indicate that the probabilities provided by our model are consistent and well-calibrated. 
We have also explored more complex models, using the same covariates, and there are not significant improvements over {\method} that justify the use of a more complex probability model over a simple and interpretable one.  
We would like to reiterate that our study does not aim at discrediting existing win probability models, but rather explore the ability of simple and interpretable models achieving similar performance to more complicated ones. 
  
One crucial point is that similar types of models need to be re-evaluated frequently.  
The game changes rapidly both due to changes in the rules but also because of changes in players skills or even due to analytics.  
This is true not only in the NFL but in other sports/leagues as well.  
For example, see the explosion of three-point shots in basketball, or the number of NFL teams that run a pass-first offense. 
Similar changes can have an implication on how {\em safe} a score differential of $x$ points is, since teams can cover the difference faster.   
For example, this can manifest itself into different coefficient for the interaction term between time lapsed and score differential.  
Hence, the details of the model can also change due to these changes.  

Finally, win probability models, while currently are mainly used from media in order to enhance sports storytelling, can form a central component for the evaluation of NFL players - of course NFL teams might be already doing this, they do so - understandably - in a proprietary manner. 
In particular, the added win probability from each player can form the dependent variable in a (adjusted) plus-minus type of regression. 
Nevertheless, the latter is a very challenging technical problem itself, given the severe co-linearities that can appear due to high overlap between the personnel of different snaps. 
In the future we plan to explore similar app 
 
 
\small

\bibliography{main-abbrv}
\bibliographystyle{abbrv}

\vspace{0.5in}

{\bf Appendix A. Calculating Team Ratings}

\vspace{0.1in}

In order to capture the opposing teams strength differential we first obtain a team rating $\rating_i$ for every team $i$, which expresses how many points better (or worse) than an average team, team $i$ is. 
Using these ratings we can express the (pregame) expected point margin between team $i$ and $j$ as $\rating_i - \rating_j$. 
We can also incorporate a home edge point $\homedge$ and express the expected point margin (assuming $i$ is the home team): $\homedge + \rating_i - \rating_j$. 
The latter will be the teams rating differential input for {\method}. 
These team ratings are updated after after every week's games. 
In order to obtain ratings for week $\week$, $\rating_i(\week)$ we will use the results from all $N$ matchups up to week $\week-1$. 
In particular, with with $\home_m$ and $\visiting_m$ being the home and visiting teams of matchup $m$, and $\margin_m$ being the win margin for the home team in matchup $m$, the solution to the following  optimization problem provides us with the team ratings:

\begin{equation*}
\begin{aligned}
& \underset{\homedge,\mathbf{\rating}}{\text{minimize}}
& & \sum_{m=1}^N (\margin_m - (\homedge + \rating_{\home_m} - \rating_{\visiting_m}))^2 \\
& \text{subject to}
& & \sum_{i=1}^{32} \rating_i = 0
\end{aligned}
\end{equation*}

One of the problems with the above optimization is the fact that for the first few weeks of the season the ratings obtained can be fairly noisy (and for week 1 there are no observations to begin with!). 
To tackle this issue we obtain pre-season ratings $\prertg_i$ for every team $i$. 
In order to obtain the pre-season ratings we make use of the expected number of wins for every team before the season begins. 
This information is available from betting sites in the form of win totals betting lines $\lambda_i$. 
As aforementioned, given team ratings $\prertg_i$ and $\prertg_j$ (and home edge), the distribution of the final point margin follows a normal distribution with mean $\homedge + \prertg_i - \prertg_j$ and standard deviation of approximately 14 points \cite{stern91}. 
Thus, the win probability for a $p$-points favorite is equal to $\Phi(\dfrac{p}{14})$, where $\Phi()$ is the cumulative distribution function for the standard normal distribution. 
Furthermore, with $\Pr[i \succ j]$ being the probability that team $i$ wins team $j$, the expected number of wins for team $i$ can be expressed as $\mathbb{E}[W_i] = \sum_{j=1}^{16} \Pr[i \succ j]$.  
We can then obtain the pre-season ratings by solving the following optimization: 

\begin{equation*}
\begin{aligned}
& \underset{\homedge,\mathbf{P}}{\text{minimize}}
& & \sum_{m=1}^{32} (\lambda_i- \mathbb{E}[W_i])^2 \\
& \text{subject to}
& & \sum_{i=1}^{32} \prertg_i = 0
\end{aligned}
\end{equation*}

After solving the two above optimization problems every team $i$ will be associated with a pre-season rating $\prertg_i$ and a season rating $\rating_i(\week)$ after week $\week - 1$.  
Pre-season ratings are our best guess for a teams strength during the beginning of the season. 
However, its impact should be reduced the more observations we obtain from the season.  
Therefore, the full rating of team $i$ after week $\week - 1$ is given by:

\begin{equation}
\rtg_i(\week) = \gamma \cdot \prertg_i + (1-\gamma)\cdot \rating_i(\week)
\label{eq:rtg}
\end{equation}
where $\gamma = \max(1-\dfrac{(\week-1)}{10},0)$, i.e., every week the impact of  the pre-seasons rating reduces linearly, and completely vanishes after week 10, when we have {\em enough} observations that can {\em over-write} the pre-season ratings.  
It should be evident that there are various other ways for someone to obtain team ratings, either by adjusting the above optimizations (e.g., by adding regularization terms, changing the cost function to the absolute value instead of the square etc.) or by considering a completely new approach (e.g., win-loss percentage, Bradley-Terry ranking \cite{agresti2003categorical}, network-based ranking \cite{pelechrinis2016sportsnetrank} etc.). 
We make use of this simple regression-based rating - among other reasons, such as its simplicity and accuracy - because it is the basis for calculating betting lines.  
Therefore, one could just incorporate the betting lines in our win probability model as is without having to calculate new ratings.

\end{document}

%% file: macros.tex
\usepackage{helvet}
\usepackage{courier}
\usepackage{epsf}
\usepackage{epsfig}
\usepackage{mathrsfs}
\usepackage{subfigure}
\usepackage{graphics}
\usepackage{balance}
\usepackage{graphicx}
\usepackage{balance}
\usepackage{epstopdf}
\usepackage{pifont}

\usepackage{amsfonts}

\usepackage{pifont}

\usepackage{epstopdf}
\usepackage{etoolbox}
\newcommand{\method}{{\tt iWinRNFL}}
\newcommand{\brier}{\beta}

\newcommand{\rating}{\mathcal{R}}
\newcommand{\homedge}{h}
\newcommand{\week}{w}
\newcommand{\home}{\mathbb{H}}
\newcommand{\visiting}{\mathbb{V}}
\newcommand{\margin}{\mu}
\newcommand{\prertg}{\rho}
\newcommand{\rtg}{\tilde{\mathcal{R}}}